\documentclass[12pt,russian]{article}
\usepackage[cp1251]{inputenc}
\usepackage[T2A]{fontenc}
\usepackage[english,russian]{babel}

\usepackage{epsfig}

\begin{document}
	
\righthyphenmin=2
	
\selectlanguage{russian}
	
\setcounter{page}{3}      
	
\newcounter{ris}
\setcounter{ris}{0}      
\newcommand{\z}{\par\refstepcounter{ris}
			{Figure  \arabic{ris}.}}

\begin{center}
	{\large{\bf{Influence of local and global solar magnetic fields on medium-term 
			forecasting of the number and parameters of coronal mass ejections}}}
\end{center}
	
\begin{center}
	{\large Irina A. Bilenko}
\end{center}
	
\begin{center}	
{\it  Sternberg Astronomical Institute of the Moscow M.V. Lomonosov State University, 
	Universitetsky pr.13, Moscow 119992, Russia, e-mail: bilenko@sai.msu.ru}
\end{center}

\section*{Abstract}
It has been shown that both the frequency and parameters of coronal
mass ejections (CMEs) increased as solar cycles 23-25 approached their 
maxima and decreased toward their minima.
However, significant differences were identified in the cyclic evolution 
of strong versus weak CMEs.
Unlike weak CMEs, powerful, high-speed CMEs, characterized by large 
angular widths and substantial mass and energy, follow the cyclic evolution 
of active regions.
During periods when the sector structure of the global magnetic field (GMF) 
dominates, the latitudinal distribution of CMEs becomes more uniform, 
and their frequency and parameter values increase, particularly during the
declining phase of solar activity.
A method based on combining GMF parameter data and Wolf numbers has 
been proposed for the medium-term forecasting of the frequency and 
parameters of strong and weak CMEs in cycle 26.
A comparison of predicted and observed CME parameters for cycle 25 
shows good agreement.

\vspace{0.5 cm}
{\noindent \bf Key words.} Sun: solar cycle - Sun: magnetic fields - Sun: 
coronal mass ejections - Sun: solar activity forecast

\section{Introduction}
\label{intro}

Coronal mass ejections (CMEs) are among the most striking and 
significant phenomena of solar activity. 
CMEs play a significant role in the formation of space weather and 
influence geomagnetic activity (Brueckner et al. 1998; Schwenn 2006; 
Shanmugaraju et al. 2015; Lawrance et al. 2016). 
This underscores the importance of CME forecasting.

A large number of studies have been devoted to CME research.
However, the nature of this phenomenon and the causes of CMEs remain unclear.
Moreover, some results are inconsistent and often contradictory. 
Predicting CMEs remains an unsolved problem, as neither the patterns of their 
formation nor their physical relationship with other solar activity phenomena 
are precisely known. At present, there is no complete clear understanding 
of the role of the influence of magnetic fields of different 
scales on the conditions of formation and cycle variations in the frequency of 
CME occurrence  and changes in their parameters during solar activity cycles.

Currently, considerable attention is being paid to operational CME forecasting.
Forecasting individual CME events and developing operational 
forecasts of their impact on space weather and geomagnetic activity requires 
a detailed study of specific events, their precursors, the state and evolution of 
photospheric magnetic fields, and CME-related 
phenomena, such as flares (Mahrous et al. 2009; Compagnino et al. 2017), 
filament/prominence eruptions (Filippov and Koutchmy 2008), 
streamers (Floyd et al. 2014), radiance variations in different bands, etc.

Modeling CME geoeffectiveness forecasts, often using state-of-the-art 
methods (Pricopi et al. 2022), is receiving considerable attention.
Recently, significant progress has been made in developing and 
constructing various models to predict the arrival time and parameters 
of disturbances caused by CMEs  (Zhao and Dryer 2014; Shlyk et al. 2023).

At the same time, medium-term, cycle-scale forecasts of the number 
and parameters of CMEs are practically nonexistent.
It is believed that the connection between the sunspot activities of the 
Sun and CMEs has already been well established (Hildner et al. 1976; 
Webb 1991; Cremades and Cyr 2007).
Only some type of CME forecasts were made based on 
active region (AR) data.
Shanmugaraju et al. (2021) conducted a study of the correlation
of yearly mean sunspot numbers in cycles 23 and 24
with the yearly occurrence rates of halo and radio loud CMEs. 
They found linear relations between them and it was used to predict 
the occurrence rates of halo and radio loud CMEs in cycle 25.
They considered two options.
For spot number equals to 109 they found
the occurrence rate of radio loud CMEs, halo CMEs, 
front-side radio loud CMEs, and front-side halo CMEs as 37, 43, 30, and 27, 
and for spot number equals to 139 -- 47, 53, 38, and 34, respectively.

In the work of Gopalswamy et al. (2023), the authors determined the power 
of cycle 25 based on the data of HALO-type CMEs in cycles 23 and 24, 
assuming that the halo CME rate provides an assessment of
the cycle strength as a weak cycle makes more halos due to the back reaction
of the weakened heliosphere on CMEs. According to their result, cycle 
25 should have been at the level of or slightly higher than cycle 24.

However, cyclical variations in the number of CMEs and their parameters
differ significantly from variations in ARs.
For example, in cycle 23, changes in CME activity followed sunspot
activity with a lag of six months to a year (Robbrecht et al. 2009; Bilenko 2014). 
Cycle variations in CME and AR latitude locations also 
differ significantly (Bilenko 2014, 2020).

Ivanov et al. (1997, 1999) and Ivanov and Obridko (2014) demonstrated that the 
large-scale structure of solar magnetic fields plays a major role in
organizing all solar activity phenomena.
Evolutionary changes in large-scale magnetic fields, which are one of the GMF 
parameters, have a significant impact on the occurrence, propagation, and 
parameters of CMEs (Ivanov et al. 1999; Fainshtein and Ivanov 2010;
Bilenko 2012, 2014, 2020).
CMEs, like many other solar activity phenomena, are controlled 
by the evolution of the solar GMF (Webb and Howard 2012).
The parameters of CMEs depend on the polar magnetic field strength and its cyclic 
variations (Petrie 2013). Bilenko (2020) identified differences in the 
behavior of CMEs with varying parameters during solar activity cycles, 
depending on the strength of the photospheric magnetic field.

All this points to the need for a detailed study of the cycle variations 
of CMEs in order to improve the quality and reliability of their forecasting.

\section{Data}   
\label{data}

For the analysis and prediction of CMEs, CME parameter data provided 
by the CDAW catalog (Gopalswamy 2009), obtained on the 
SOHO spacecraft using the LASCO instrument for cycles 23-25 were used.

Wolf number data from the WDS SILSO database (Clette and Lef\`evre 2015) 
were used to predict solar activity cycles and comparison of AR and CME
parameter cycle variations.

To account for the influence of the Sun's GMF on the number
and parameters of CMEs, data from the Wilcox Solar Observatory (WSO), 
regularly provided since 1976, were used (Hoeksema 1984; Hoeksema 1986).

\section{CME parameters in cycles 23\,--\,25}   
\label{cmes}

\begin{figure}     
	\begin{center}
		\epsfig{file=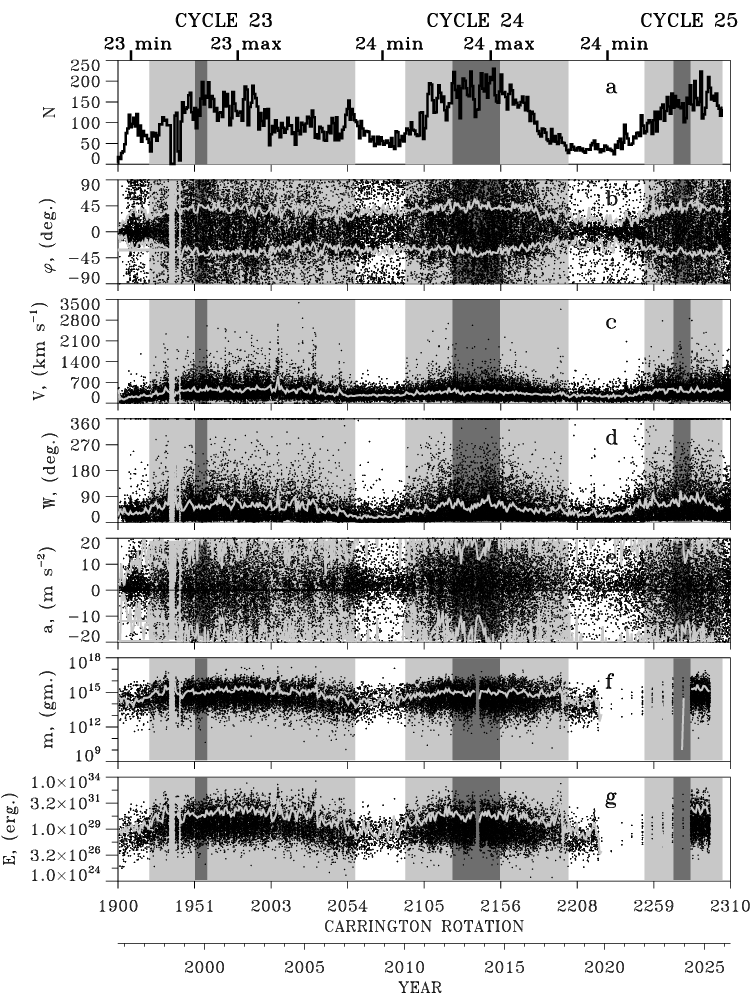,height=19cm}
	\end{center}
	\z{ CME number in a CR (a) and  CME parameters:
		(b) location in latitude;
		velocity (c) and  angular width (d) in sky plane;
		(e) acceleration;
		(f) mass;
		(g) kinetic energy.
		The light gray background areas correspond to periods of the GMF sector 
		structure domination, while the dark-gray areas represent the times 
		of polarity reversal in each cycle.}
	\label{cme_all} 
\end{figure}

\begin{figure}    
	\begin{center}
		\epsfig{file=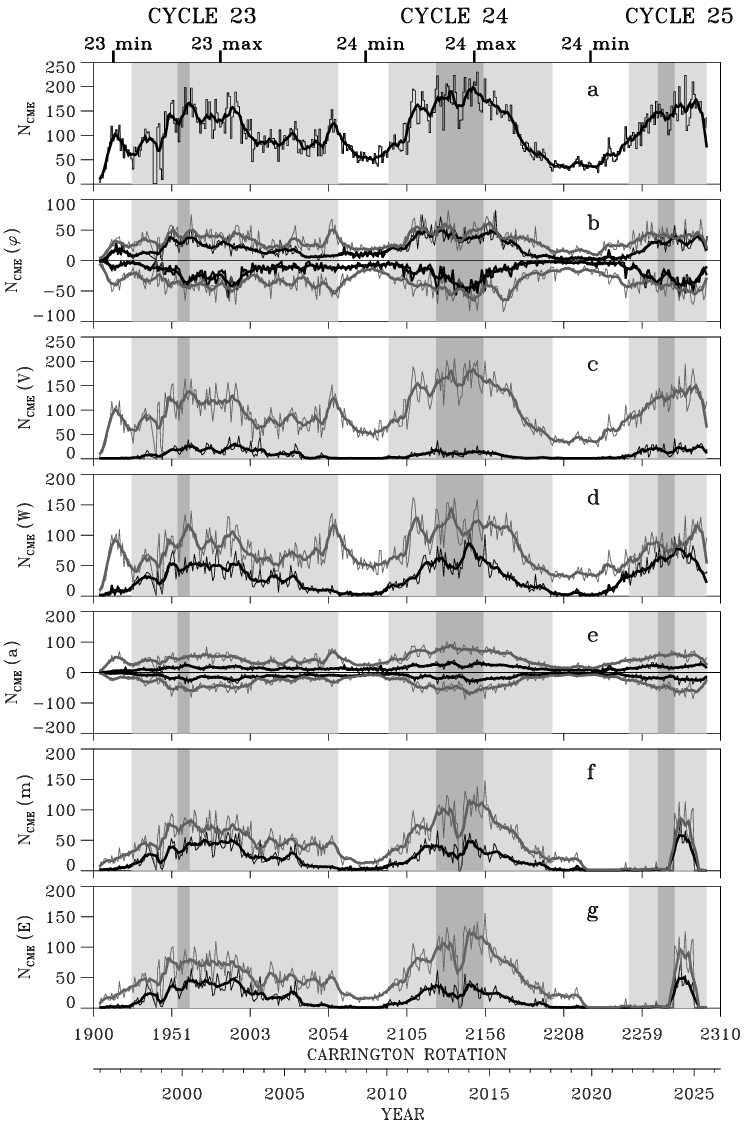,height=19cm}
	\end{center}
	\z{ (a) CME number in a CR. 
		 The number of CMEs with parameters, which are greater or 
		 equal (black lines) and smaller then (gray lines) threshold values:
		(b) location in latitude;
		velocity (c) and  angular width (d) in sky plane;
		(e) acceleration;
		(f) mass;
		(g) kinetic energy.
		Thin lines denote CR-averaged values, while thick lines 7 CR-averaged ones.
		The description of the background shading is the same as in Figure~\ref{cme_all}.}
	\label{cme_cr_lim}
\end{figure}

\begin{figure}    
	\begin{center}
		\epsfig{file=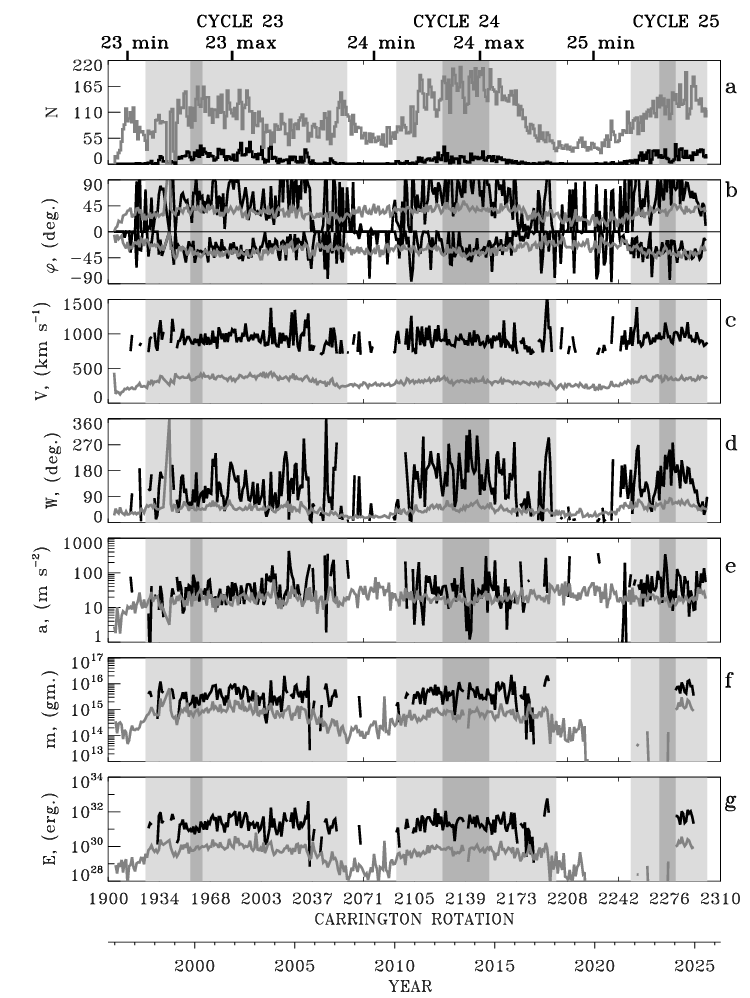,height=19cm}
	\end{center}
	\z{ CME number in a CR (a) and parameters, which 
		are greater or equal (black lines) and smaller 
		then (gray lines) threshold values:
		(b) location in latitude;
		velocity (c) and  angular width (d) in sky plane;
		(e) acceleration;
		(f) mass;
		(g) kinetic energy.
		The description of the background shading is the same as in Figure~\ref{cme_all}.}
	\label{cme_vg_vs}
\end{figure}

Figure~\ref{cme_all} shows variations in the number 
and parameters of CMEs in cycles 23-25 based on data from the CDAW catalog.
Each point on the graphs in Figures~\ref{cme_all}b-g corresponds 
to a specific CME event, and the light-gray lines show changes in the 
7 CR (Carrington rotation) average values of each CME parameter. 
The light gray background areas correspond to periods of the GMF sector 
structure domination, while the dark-gray ones represent the times 
of polarity reversal at solar activity maxima in each cycle.

The time of GMF sectorial structures domination was determined
based on data from the decomposition
of solar magnetic fields into spherical harmonics.
The solar magnetic field can be represented
as a function of distance, latitude, and longitude coordinates  ($r, \theta, \phi$):
\begin{equation}
	\psi(r,\theta,\phi)=R \sum_{n=1}^N \sum_{m=0}^n
	\left(\frac{R}{r}\right)^{n+1} [g_n^m \, {\rm cos}\left(m\phi\right)\, + \, h_n^m \, {\rm 
		sin}\left(m\phi\right)] \, P_n^m(\theta), 
	\label{r_theta_psi}
\end{equation}
\noindent where: $P_{n}^m(\theta)$ are the associated Legendre polynomials, 
and $N$ is the number of harmonics (Altschuler et al. 1975, 1977).
The coefficients $g_n^m$ and $h_n^m$ are determined
by a method based on the orthogonality of spherical harmonics (Hoeksema 1984).
$g_n^m$ and $h_n^m$ are calculated using a least-squares fit to the observed
radial component of the photospheric magnetic fields
under the assumption of a potential magnetic field.
The equation~\ref{r_theta_psi} describes the solar magnetic field
as a sum of individual functions.
Each harmonic, defined by its indices ($n$ and $m$),
corresponds to the contribution of certain components to the overall magnetic field 
distribution. Based on the coefficients
$n$ and $m$ obtained from the equation~\ref{r_theta_psi}, the power spectrum of the 
various harmonic components can be calculated (Levine 1977): 
\begin{equation}
	S_n=\sum_{m=0}^n [(g_n^m)^2 + (h_n^m)^2].
	\label{g1_f_spectr}
\end{equation}

For $n = m$, the functions are called sectorial and correspond to the sectorial 
structures of the GMF, which are alternating meridional regions of 
positive- and negative-polarity magnetic fields.
The evolution of individual harmonic components was examined 
in detail in the papers: Obridko (2023) and Obridko (2024).

The moments of polarity reversal at the north and south poles and their total 
duration in each cycle, shown in Figures~\ref{cme_all} -- \ref{cme_vg_vs}, were 
determined based on measurements of polar magnetic fields at the WSO 
observatory (Bilenko 2026).

It follows from Figure ~\ref{cme_all} that both the number and parameters 
of CMEs, in general, vary in accordance with cycle variations
in the number of ARs characterized by Wolf numbers,
as has been repeatedly noted by a number of researchers.
Toward cycle maxima, both the number of CMEs and the values of their 
parameters increase. The latitude distribution of CMEs becomes more
uniform, and CMEs are observed at all latitudes in the northern and
southern hemispheres (Bilenko 2020). This corresponds to a period
of the GMF sectorial structure domination, determined by the increasing 
influence of sectorial harmonics.

However, it should also be noted that CMEs with low parameter values
are observed all the time in all cycles, regardless of the cycle phase.
An examination of cyclic variations in CME parameters depending on their 
magnitude in cycles 23 and 24 revealed significant differences
in the behavior of strong and weak CMEs, respectively, with high
and low velocities, large and small  angular width,
mass, and energy during solar activity cycles (Bilenko 2020).
According to Bilenko (2020), CMEs with parameters above and below 
the corresponding threshold values were separately identified,
which were:
$\phi$=40$^{\circ}$,
V=700 km/s,
W=60$^{\circ}$,
a=20 m/s$^2$,
m=10$^{15}$ g, and
E =10$^{30}$ erg. (Figure~\ref{cme_cr_lim}).

Figure~\ref{cme_cr_lim} shows cycle variations
in the CME number  (Figure ~\ref{cme_cr_lim}a) 
and CME parameters whose values are equal to and above, or below
the corresponding threshold value (Figure ~\ref{cme_cr_lim}b-g) in cycles 23-25.
From Figure~\ref{cme_cr_lim} it follows that the nature of the cycle evolution 
of the number of CMEs with parameters equal to or greater than the 
threshold values differs significantly from the variations in the 
number of CMEs with parameters below the threshold values. 
The number of CMEs with parameters above the threshold
demonstrates cycle changes characteristic of Wolf numbers,
while the behavior of the number of CMEs with parameters below the threshold
does not reveal such a relationship. The number of weak CMEs increases 
significantly in the low cycle 24, which is a consequence of the decrease
in the strength in both the polar (Petrie 2013, 2015) and non-polar
(Bilenko 2020) magnetic fields, which are the visible manifestation 
of cyclic variations in the poloidal and toroidal components of the GMF.

It should be noted that, regarding mass and energy parameters, 
the cycle variations of CMEs with parameter values both above 
and below the threshold levels are nearly identical.
This is due to the fact that the CDAW catalog does not contain
mass and energy data for weak CMEs with angular width
less than 20 degrees. 
Furthermore, the mass and energy parameters are calculated 
values and are often entered into the CDAW catalog with a significant 
time lag, resulting in a lack of mass and energy data for extended periods
of time (Figures~\ref{cme_all} and \ref{cme_vg_vs}).

Figure~\ref{cme_vg_vs} shows variations in the parameters of CMEs that 
have velocities equal to or higher (black lines) and lower (gray lines)
than the threshold values.
These dependencies confirm that CMEs with high
velocities, on average,  have also large angular width,
acceleration, mass, and energy, i. e., these are primarily powerful CMEs.
Figure~\ref{cme_vg_vs}e shows values only for CMEs with positive 
acceleration, since for the acceleration values averaged over a CR,
the dependencies for positive and negative accelerated
CMEs are practically identical.

An analysis of Figure~\ref{cme_vg_vs} reveals that the time-dependence 
curves of CME parameters for powerful events are asymmetric, showing 
higher values during the decay phases (Figure~\ref{cme_vg_vs}b-g). 
This can be explained by the fact that the sectorial structure of the GMF is 
more clearly manifested during the decay phases. Furthermore, many 
large ARs form during these phases. They are characterized by 
heightened flare activity, including a high number of intense flares, 
which leads to an increase in the occurrence of powerful CMEs.
The increase in the number of weak CMEs during the declining phases 
of solar cycles 23 and 24 (Figure~\ref{cme_vg_vs}a), most notably 
during the declining phase of cycle 23, is associated with processes 
of GMF restructuring, during which the number of weak 
CMEs rises sharply (Bilenko 2014).

\section{Solar cycle 26  Wolf number and CME forecast}   
\label{cycle_26_forecast}

\begin{figure}    
	\begin{center}
		\epsfig{file=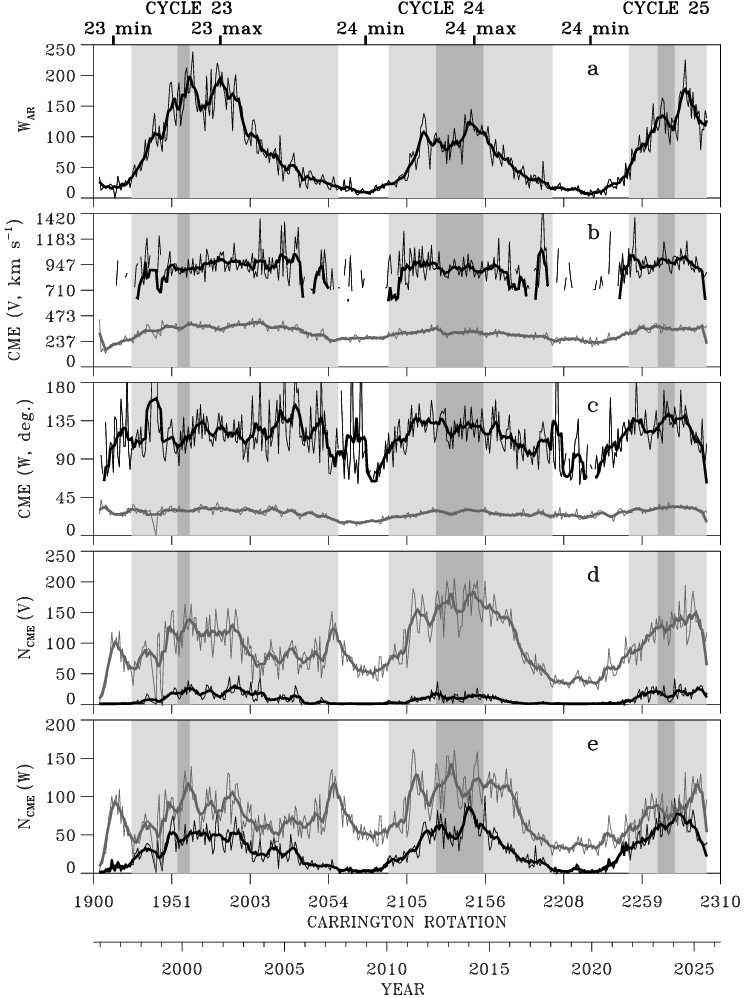,height=19cm}
	\end{center}
	\z{ (a) Wolf numbers;
		velocity (b) and  angular width (c) of CMEs in sky plane;
		number of CMEs with velocity (d) and  angular width (e) in sky plane 
		for CMEs with parameters greater (black lines) and smaller
		then (gray lines) values.
		Thin lines correspond to values averaged over 1 CR,
		and thick lines to values averaged over 7 CRs.
		The description of the background shading is the same as in Figure~\ref{cme_all}.}
	\label{cme_V_W_N}
\end{figure}

For forecasting purposes, the parameters of  CME velocity and angular width 
were selected, since these quantities are directly measured 
and are most fully presented in the CDAW catalog.
Figure~\ref{cme_V_W_N} compares the changes in the Wolf numbers
(Figure~\ref{cme_V_W_N}a), the CR averaged values of CME velocity 
and angular width (Figures~\ref{cme_V_W_N}b, c), and
their numbers (Figures~\ref{cme_V_W_N}d, e) for CMEs 
with parameter values equal to or above (black lines) and below (gray lines)
than the corresponding threshold values.
From Figures~\ref{cme_all} -- \ref{cme_vg_vs} it follows that elevated
values of the CME parameters and their numbers are observed during 
periods  of the GMF sectorial structures domination in each cycle.

Changes in the Wolf numbers show fairly distinct, repeating cyclic variations 
with a period of approximately 11 years for over 100 years.
As noted in the section~\ref{cmes}, these variations
are clearly visible in the cycle changes of powerful CMEs with high
velocities and large angular width, measured in the sky plane, with increased 
mass and energy.
The correlations coefficients between Wolf numbers and the number of CMEs,   
for CMEs  with velocities and angular width equal to or above the 
threshold values. are approximately 0.702 and 0.7, respectively.

\begin{figure}    
	\begin{center}
		\epsfig{file=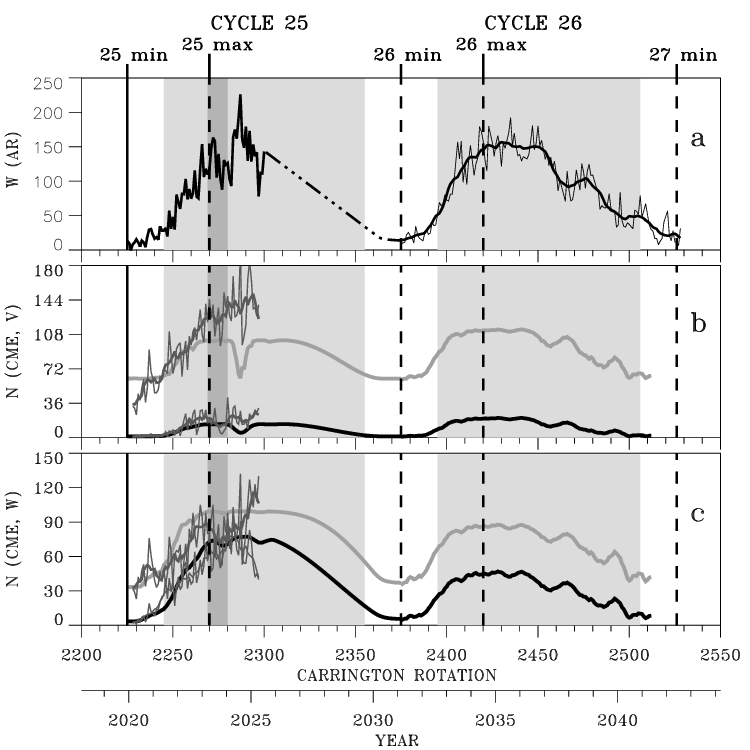,width=15cm,height=17cm}
	\end{center}
	\z{ Actual cycle 25 and predicted cycle 25 and 26 Wolf number and 
		CME parameters.}
	\label{cycle_26_shema}
\end{figure}

Figure ~\ref{cycle_26_shema}a shows the changes in Wolf numbers in
the current  cycle 25 and their forecast for cycle 26.

An investigation of positive- and negative-polarity magnetic fields
has revealed the presence of a meridional circulation of the GMF that 
differs from the circulation characteristic of ARs (Bilenko 2024, 2026).
The timing and duration of polarity reversals also do not correspond 
to any specific characteristics of Wolf numbers (Bilenko 2026).
The approximation of the meridional circulation of large-scale
positive- and negative-polarity magnetic fields 
by a sum of sinusoids (Bilenko 2026) allows to predict
the timing of polarity reversals (magnetic field sign changes at the solar poles).
Equation for approximating the magnetic field flow of positive polarity:
\begin{eqnarray} 
	L{^\circ}_{\rm pos}(\rm CR) = -2.8 - 26 \times {\rm cos}\left(\frac{t}{45.9} 
	- 0.87\right) + 5.3 \times {\rm cos}\left(\frac{t}{17.05} + 0.87\right)
	\label{eq_pos}	
\end{eqnarray}
and that for negative-polarity magnetic field flow:
\begin{eqnarray}
	L{^\circ}_{\rm neg}(\rm CR) = -3 - 24 \times {\rm sin}\left(\frac{t}{46.2} 
	- 1.97\right) + 4.7 \times 	{\rm sin}\left(\frac{t}{17.25}  + 0.5\right)
	\label{eq_neg}	
\end{eqnarray}
\noindent where $L$ is latitude in degrees and $t$ is time in CR number
(Bilenko 2026).

These equations make it possible to determine the approximate timing 
of the polarity reversals.
A comparison of the polarity reversal times determined by
equations ~\ref{eq_pos} and ~\ref{eq_neg} with the peak times 
from cycles 11 to 25 
obtained by WDS SILSO shows good agreement (Bilenko 2026).
Using equations ~\ref{eq_pos} and ~\ref{eq_neg}, the approximate dates of 
the polarity reversals for cycles 26 and 27 were determined, which were 
CR 2420 (02.07.2034-29.07.2034)  and CR 2571 (11.10.2045-07.11.2045), 
respectively.
Considering also that polarity reversals always occur during the phase 
of  solar activity maximum, the obtained values can be regarded as 
approximate times of  activity maxima in cycles 26 and 27.
The spread between the polarity reversal times obtained using
equations ~\ref{eq_pos} and \ref{eq_neg} and the cycle maximum 
dates provided by the SILSO WDS ranges from 2 to 27 CRs (Bilenko 2026).
Therefore, the maximum period of cycle 26 could range 
from CR 2406.5 to CR 2433.5 (28.06.2033-06.07.2035).

It was shown that polarity reversals began 50, 44, 42, 46, and 44 CRs 
after the activity minimum dates listed in the SILSO database 
for cycles 21, 22, 23, 24, and 25 (Bilenko 2026).
The average value is 45 CRs.
Therefore, the approximate date of the cycle 26 minimum can 
be determined as 2420-45, which yields a CR of 2375 (21.02.2031-20.03.2031) $\pm$5.
The still-incomplete decline phase of the cycle 25 was artificially extended 
to this CR (Figure~\ref{cycle_26_shema}а).

To determine the duration of cycle 26, the polarity reversal and minimum
times of cycle 27 were similarly calculated.
These were CR 2571 (11.10. 2045-07.11.2045) $\pm$5 
and CR 2526  (01.06. 2042-28.06.2042) $\pm$5, respectively.
In Figure ~\ref{cycle_26_shema}, the vertical dotted lines
show the calculated times of the cycles 26 and 27 minima and
cycles 25 and 26 maxima.
According to the obtained Wolf number prediction results
for the minima of cycles 26 and 27, the duration of cycle 26
will be approximately 151 CR, or approximately 11.2838 years.

Assuming, following Ger\d{c}eker (2025),
that cycle 26 will be close in its parameters to cycle 20,
the contours of cycle 20 are shown in Figure ~\ref{cycle_26_shema}a. 
The duration of cycle 20 matches that of predicted cycle 26.

Since cycle changes in the number of CMEs with parameters above 
the threshold correspond fairly well to variations in Wolf numbers,
it is possible to make a forecast for the parameters of powerful CMEs
using the forecast data for Wolf numbers in cycle 26.
The number of CMEs with parameters above the threshold values was determined
based on the ratio of Wolf numbers and the number of CMEs in cycle 23, since
in terms of the ratio of the number of CMEs with parameters above and below 
the threshold values, cycle 24 differs sharply from both cycles 23 and 25.
Third-degree polynomial approximations were calculated for
the dependencies between Wolf numbers and the number of CMEs 
with velocities above the threshold value and that with 
angular width in cycle 23. 
The values of both the Wolf numbers and the CME parameters were 
smoothed by 7~CRs to reduce the influence of random fluctuations.
The coefficients obtained during the approximation were used in 
expressions for calculating the CME predicted numbers in cycle 26:
\begin{eqnarray} 
	N(V)  =  -0.38748  +  0.08259 \times x  + 0.00114 \times x^2 - 0.00001 \times x^3
	\label{n_vg}	
\end{eqnarray} 
\begin{eqnarray} 
	N(W) =  -2.89168  + 0.56841 \times x - 0.00245 \times x^2 + 0.00001 \times x^3
	\label{n_wg}	
\end{eqnarray} 
x - predicted Wolf numbers for cycles 25 and 26.

The obtained values of the CME number with velocities and  angular width
equal to or greater than the threshold values for cycles 25 and 26
are shown in Figures~\ref{cycle_26_shema}b and c as thick black lines.

Cyclical variations in the number of CMEs with parameters below the
threshold do not allow them to be unambiguously associated with any 
solar activity phenomenon.
Therefore, for cycles 25 and 26, their numbers were calculated based 
on the ratio of the  CME numbers with parameters above 
and below the threshold in cycle 23, also using third-degree polynomial 
approximation. 
Since sharp spikes in the number of CMEs with low parameters  
at the beginning and end of cycle 23 were absent in cycles 
24 and 25 during the corresponding cycle phases,
the time interval from CR 1933 (18.02.1998-17.03.1998)
to CR 2038 (22.12.2005-18.1.2006) of cycle 23 was 
used to determine the approximation coefficients.

Using the coefficients obtained during the approximation, the 
number of CMEs with velocity and angular width 
below the threshold  value were calculated for cycles 25 and 26:
\begin{eqnarray} 
	N(V)  =  56.39516  + 4.18909\times x - 0.06956\times x^2 - 0.00001\times x^3
	\label{n_vs}	
\end{eqnarray} 
\begin{eqnarray} 
	N(W)  =  26.80932 + 1.93448\times x - 0.01456\times x^2 + 0.00002\times x^3
	\label{n_ws}	
\end{eqnarray}

The obtained values for cycles 25 and 26 are shown in
Figures~\ref{cycle_26_shema}b and c as thick gray lines.
The thin mid-gray lines show the observed
values of the CR-averaged number of CMEs for the corresponding
parameters (Figure~\ref{cycle_26_shema}b, c).
In cycle 25, the CR-averaged observed parameter values  
 show fairly good consistency with the predicted ones  
 (Figures~\ref{cycle_26_shema}b, c).

In cycles 21-25, the sectorial structure of the GMF began to dominate 
approximately 15-25 CRs after the minimum, and its dominance ended 
also approximately 15-25 CRs before the minimum  of the next cycle, 
indicated in the WDS SILSO database.
On average, this is equal to 20 CRs.
Accordingly, the beginning and end dates of the  GMF 
sectorial structure domination  will be 
CRs 2245 (07.06.2021-04.07.2021) and 2355 (24.08.2029-20.09.2029)
in cycle 25 and CRs 2395 (19.08.2032-15.09.2032) 
and 2546 (29.11.2043-26.12.2043) in cycle 26.
During these periods of the GMF sectoral structure domination 
an increase in the CME number and parameters is expected. 
The values of the average velocity and  angular width of 
powerful CMEs over CRs  remained approximately the same 
in cycles 23 and 24. 
It is assumed that in cycle 26 their values will remain at the same level.

\section{Conclusion}  
\label{conclud}

It has been shown that CMEs with parameters greater than or equal to
$\phi$=40$^{\circ}$, V=700 km/s, W=60$^{\circ}$,
a=20 m/s$^2$, m=10$^{15}$ g, E=10$^{30}$ erg. and that with parameters
below these threshold values exhibit different dynamics during cycles 23-25.

Based on the obtained results, separate forecasting of cycle variations 
in strong and weak CMEs is proposed.
To forecast CMEs in cycle 26, data on the CME velocity and angular  width
parameters were 
selected, as they are directly measured by the SOHO spacecraft
and are most fully represented in the CDAW catalog.

Powerful, high-velocity CMEs, characterized by large angular width, 
significant mass, and high energy, follow the cycle evolution of ARs.
The number and  parameters of weak CMEs  do not reveal 
a similar relationship. The number of weak CMEs increased significantly 
in cycle 24.

Based on the meridional circulation of the GMF (Bilenko 2024, 2026),
the approximate dates of polarity reversals in cycles 26 and 27 
were determined, which were 
CR 2420 (02.07.2034-29.07.2034) and CR 2571 (11.10.2045-07.11.2045), 
respectively.
This, taking into account previously identified parameters of GMF, 
allowed to determine the approximate times of the minima of cycles 26 CR 2375 
(21.02.2031-20.03.2031) $\pm$5 
and 27 CR 2526  (01.06. 2042-28.06.2042) $\pm$5,
the duration of cycle 26 (151 CRs or 11.2838 years),
as well as the periods of the GMF sectorial structure domination in cycles 25 
from CR 2245 (07.06.2021-04.07.2021) to CR 2355 (24.08.2029-20.09.2029) 
in cycle 25 and from CR 2395 (19.08.2032-15.09.2032) to 
CR 2546  (29.11.2043-26.12.2043) in cycle 26.

CRs 2245 (07.06.2021-04.07.2021) and 2355 (24.08.2029-20.09.2029)
in cycle 25 and CRs 2395 (19.08.2032-15.09.2032) 
and 2546 (29.11.2043-26.12.2043) in cycle 26.

Assuming similarities between cycles 20 and 26 (Ger\d{c}eker 2025),
a forecast of cycle variations in Wolf numbers in cycle 26 was made.
Using these data, a forecast of the number of powerful CMEs 
with velocity and  angular width above the threshold values was made
for cycles 25 and 26 .
Based on the ratio of the numbers of weak and strong CMEs during cycle 23, 
a forecast was made for weak CMEs, characterized by velocities and 
angular widths below threshold values, for cycles 25 and 26.
A comparison of the predicted CME parameters with those observed in 
the current cycle 25 shows good agreement.

The CR-averaged velocities and angular width of strong CMEs 
remained approximately the same during periods of  the GMF sectorial
structure domination in cycles 23 and 24. It is expected that
in cycle 26, their values will remain at the same level during a similar period.

The results indicate that the method based on a combination of data on
the GMF and AR parameters  can be useful in medium-term forecasting 
of the number and parameters of different type CMEs.

\vspace{0.5cm}

{\large\bf Acknowledgements}


The study was conducted under the state assignment of Lomonosov 
Moscow State University.

This CME catalog is generated and maintained at the CDAW Data Center 
by NASA and The Catholic University of America in cooperation with the 
Naval Research Laboratory. SOHO is a project of international cooperation
 between ESA and NASA.

Wilcox Solar Observatory data used in this study was obtained via
the web site
$http://wso.stanford.edu$ at $2026:02:26\_00:54:03$ PST
courtesy of J.T. Hoeksema. The Wilcox Solar Observatory
is currently supported by NASA.

Data from WDS SILSO (Sunspot Index and Long-term Solar Observations) 
data base of the Royal Observatory of Belgium were used in the analysis 
of Wolf numbers (https://doi.org/10.24414/qnza-ac80).

\vspace{0.5cm}

{\large\bf  References}

\noindent Altschuler M.D., Trotter D.E., Newkirk G.Jr., Howard R. (1975) Solar Phys. 41, 
225

\noindent Altschuler M.D., Levine R.H., Stix M., Harvey J. (1977) Solar Phys. 51, 345

\noindent Bilenko I.A. (2012) Geomagnetism and Aeronomy 52, 1005

\noindent Bilenko I.A. (2014) Solar Phys. 289, 4209

\noindent Bilenko I.A. (2020) ApJ 889, id.1

\noindent Bilenko I.A. (2024) Solar Phys. 299, id. 103

\noindent Bilenko I.A. (2026) Solar Phys. 401, id. 42

\noindent Brueckner G.E., Delaboudiniere J.-P., Howard R.A.  et al. (1998) 
Geophys. Res. Lett. 25, 3019

\noindent Clette F. and Lef\`evre L. (2015) WDC SILSO ROB

\noindent Compagnino A., Romano P., Zuccarello F. (2017) Solar Phys. 202, id. 5

\noindent Cremades H., St. Cyr O.C. (2007) Adv. Space Res. 40, 1042

\noindent Fainshtein V.G., Ivanov E.V. (2010) Sun and Geosphere 5, 28

\noindent Filippov B., Koutchmy S. (2008) Annales Geophysicae 26, 3025

\noindent Floyd O., Lamy P., Llebaria A. (2014) Solar Phys. 289, 1313

\noindent Ger\d{c}eker K., Kilcik A., Ozguc A., Yurchyshyn V. (2025) Solar Phys. 300, id. 
169 

\noindent Gopalswamy N., Yashiro S., Michalek G., et al, (2009) EM\&P 104, 295

\noindent Gopalswamy N.,  Michalek G., Yashiro S. (2023) ApJL 952, 13 

\noindent Hildner E., Gosling J.T., MacQueen R.M. et al. (1976) 48, 127

\noindent Hoeksema J.T. (1984) Ph. D. Thesis

\noindent Hoeksema J.T., Scherrer P.H. (1986) Solar Phys. 105, 205

\noindent Ivanov E.V., Obridko V.N., Shelting B.D. (1997) ARep 41, 236

\noindent Ivanov E.V., Obridko V.N., Nepomnyashchaya E.V., Kutilina N.V. (1999) 
Solar Phys. 184, 369

\noindent Ivanov E.V., Obridko V.N. (2014) Geomagnetism and Aeronomy 54, 996

\noindent Levine R.H. (1977) Solar. Phys. 72, 277

\noindent Lawrance M.B., Shanmugaraju A., Moon Y.-J. et al. (2016) Solar Phys. 291, 
1547

\noindent  Mahrous A., Shaltout M., Beheary M.M. et al.(2009) Adv. Space Res. 43, 103

\noindent Obridko V.N., Shibalova A.S., Sokoloff D.D. (2023)  MNRAS 523, 982

\noindent Obridko V.N., Shibalova A.S., Sokoloff D.D. (2024)  MNRAS 529, 2846

\noindent Petrie G.J.D. (2013) ApJ 768, 162

\noindent Petrie G.J.D. (2015)  ApJ 812, id. 74

\noindent Pricopi A.-C., Paraschiv A.R., Diana Besliu-Ionescu D., 
Marginean A.-N. (2022) ApJ 934, id. 176

\noindent Robbrecht E., Berghmans D., Van der Linden (2009) ApJ 691, 1222

\noindent Schwenn R. (2006) Living Rev. Sol. Phys. 3, id. 2

\noindent Shanmugaraju A.,  Pappa Kalaivani P.,  Moon Y.-J., Prakash O. (2021)
Solar Phys.  296, 75

\noindent Shlyk N.S.,  Belov A.V., Abunina M.A.,  Abunin A.A.  (2023)
Geomagn. Aeron. 63, 564 

\noindent Webb  D.F. (1991) Adv. Space Res.  22, 37

\noindent Webb D.F., Howard T.A. (2012)  Living Rev. Sol. Phys. 9, id. 3

\noindent Zhao X., Dryer M. (2014) Space Weather 12, 448

\end{document}